\documentclass[aps,prd,preprintnumbers,twocolumn,groupedaddress,nofootinbib]{revtex4}

\pdfoutput=1

\usepackage{graphicx, color}
\usepackage{latexsym}
\usepackage{amsfonts}
\usepackage{amssymb}
\usepackage{amsmath}
\usepackage{slashed}
\usepackage{array}
\usepackage{feynmp}
\usepackage{url}
\usepackage{ulem}

\begin{document}
\title{Leptophobic $Z'$ Boson and Parity-Violating $eD$  Scattering}

\author{Mart\'in Gonz\'alez-Alonso$^{1}$ and Michael J. Ramsey-Musolf$^{1,2}$}
\affiliation{$^1$Department of Physics, University of Wisconsin, Madison, WI 53706, USA}
\affiliation{$^2$Kellogg Radiation Laboratory, California Institute of Technology, Pasadena, CA 91125, USA}

\preprint{NPAC-12-18}
\date{\today}

\begin{abstract}
We study the impact of a leptophobic $Z'$ gauge boson on the $C_{1q}$ and $C_{2q}$ parameters that describe the low-energy, parity-violating electron-quark neutral current interaction. We complement  previous work by including the penguin-like vertex corrections, thereby completing the analysis of one-loop calculation up to ${\cal O} (m^2_{q'} / M^2_{Z'})$ terms. We analyze the sensitivity of these probes to the different couplings $Z' \bar{u} q$ ($q=u,c,t$) and $Z' \bar{d} q$ ($q=d,s,b$), in a model-independent way that can be applied to any specific $Z'$ scenario. We show that constraints from neutral kaon and heavy flavor studies preclude significant contributions from flavor non-diagonal couplings except for those involving top quarks.
We apply our results to a light $Z'$ with flavor diagonal couplings to up or down quarks, a scenario proposed in the literature to explained the CDF $W$ plus di-jet anomaly. We find that such a particle would not affect the $C_{1q}$ coefficients, but it would have a sizable impact on $C_{2q}$ couplings that can be probed by future measurements of parity-violating deep inelastic scattering of polarized electrons off of deuterium. 
\end{abstract}

\maketitle


Despite the impressive success of the Standard Model (SM), it is widely believed that it is not the final theory and that some extension is needed. One of the simplest modifications is the addition of an extra $U(1)$ gauge group, with the consequent addition of a $Z'$ gauge boson. This extension is present in many scenarios for physics beyond the Standard Model (BSM), and the details of the $Z'$ interactions with fermions are completely model-dependent. 

The possibility of a leptophobic $Z'$ boson, both with diagonal and non-diagonal couplings in flavor space, has received some attention recently in the literature as a possible explanation for some experimental anomalies \cite{Yu:2011cw,Buckley:2011mm,Buckley:2011vc,Buckley:2011vs,Cheung:2011zt}. Interestingly, this kind of $Z'$ scenario is difficult to probe with colliders due to the QCD background, and the limits on them are not strong.

In this paper we analyze the possibility of indirectly probing these models using parity-violating (PV) deep inelastic scattering (DIS) of electrons off of deuterium. The measurement of this PV asymmetry has been carried out recently at Jefferson Lab (JLab) \cite{Subedi:2011zz}, and more precise measurements are planned for the JLab 12 GeV program \cite{Dudek:2012vr,SOLID} and possibly for a future Electron Ion Collider (EIC) \cite{Boer:2011fh}.

The low-energy effective Lagrangian describing the PV interaction of an electron and a light quark $q=u,d$ is
\begin{equation}
{\cal L}_{\rm PV} = \frac{G_F}{\sqrt{2}} \sum_q \left[ C_{1q} (\bar{e} \gamma^\mu \gamma_5 e)( \bar{q}\gamma_\mu q )+ C_{2q} (\bar{e} \gamma^\mu e)( \bar{q}\gamma_\mu \gamma_5 q ) \right],
\label{eq:efflag}
\end{equation} 
where $G_F$ is the Fermi constant that can be determined from the muon lifetime. At tree-level in the SM, these couplings are given by
\begin{eqnarray}
\label{C12tree}
C_{1q} & = & - I_3^q + 2 Q_q \sin^2 \theta_{\mathrm{W}}~, \label{eq:C1} \\ 
C_{2q} & = & I_3^q \left( -1 +4 \sin^2\theta_{\mathrm{W}}\right)~, \label{eq:C2}
\end{eqnarray}
where $Q_q$ and $I_3^q$ are the electric charge and the third component of weak isospin for the quark $q$, and $\theta_{\mathrm{W}}$ is the weak mixing angle. The ${\cal O}(\alpha)$ electroweak radiative corrections can be found in Refs.~\cite{Marciano:1982mm,Musolf:1990ts,Beringer:1900zz}, and the theoretical uncertainty in $C_{1q,2q}$ is below the percent level. The coefficients $C_{1q}$ have been extracted with high precision \cite{Young:2007zs} from atomic PV in Cesium \cite{Wood:1997zq,Porsev:2009pr} and will be determined also through PV elastic $ep$ scattering \cite{QWeak}. On the other hand, the current experimental determinations of the $C_{2q}$ coefficients are much less precise (cf. Ref.~\cite{Beringer:1900zz}), although this situation will change in the near future due to PV-DIS studies. We will see in this paper how $C_{2q}$ are sensitive to some BSM models that do not affect the $C_{1q}$ coefficients, showing their complementarity as BSM precision probes.

More specifically, the parity-violating $eD$ asymmetry depends on the $C_{iq}$ coefficients as follows
\begin{eqnarray}
\label{CG-modify}
A_{PV}^{eD} &=& - \frac{G_F Q^2}{2\sqrt{2}\pi \alpha}\frac{9}{10}\Big [\tilde{a}_1  + \tilde{a}_2 \frac{1-(1-y)^2}{1+(1-y)^2}\Big ]~,\\
\tilde{a}_1 &=& -\frac{2}{3}\left(2C_{1u}-C_{1d}\right) \big [1 + R_1  \big ], \\
\tilde{a}_2 &=& -\frac{2}{3}\left(2C_{2u}-C_{2d}\right) \big [1  + R_2 \big ],
\end{eqnarray}
where $Q^2 = |{\vec q}|^2-q_0^2$ is the four-momentum transfer squared, $y$ is the fractional energy transfer from the electron to the hadrons and $R_{1,2}$ include different hadronic corrections (see Ref.~\cite{Mantry:2010ki} for more details and references). The SOLID experiment at JLab will measure this asymmetry with a statistical precision better than 0.5\% for a number of bins with $0.3<x<0.75$, $4<Q^2<10$ (GeV/c)$^2$, and $y\sim1$ \cite{SOLID}. SOLID has a significant sensitivity to the $C_{2q}$ coefficients due to the large value of $y$, and is projected to determine $2C_{2u} - C_{2d}$ with an error $ \pm 0.0083$ \cite{KK}, which could be improved by an EIC measurement.

\begin{figure}[h!]
\begin{minipage}[t]{.35\linewidth}\centering
\centerline{\includegraphics[width=3.5cm]{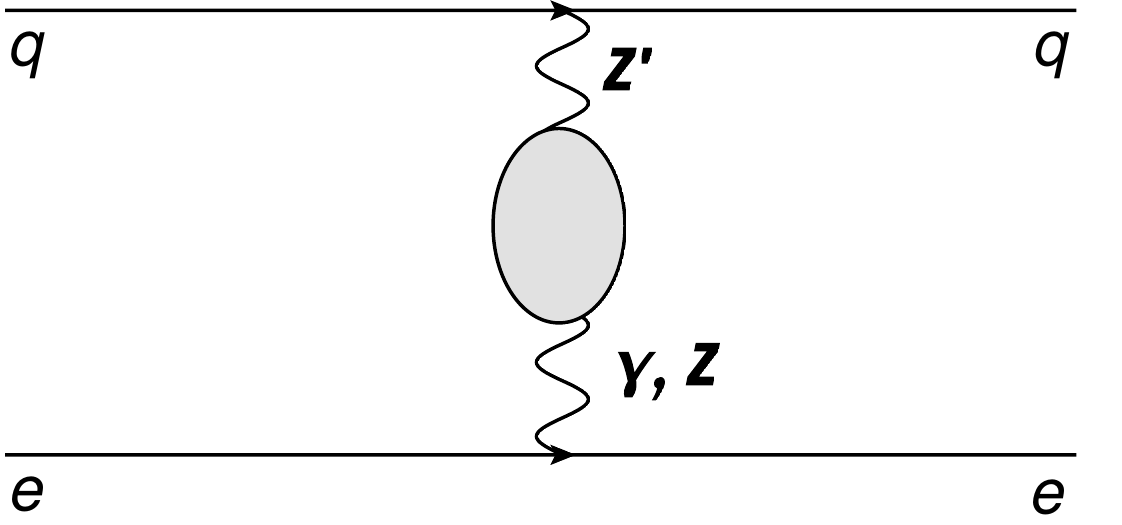}}
\end{minipage}
\hspace{1cm}
\begin{minipage}[t]{.35\linewidth}\centering
\centerline{\includegraphics[width=3.5cm]{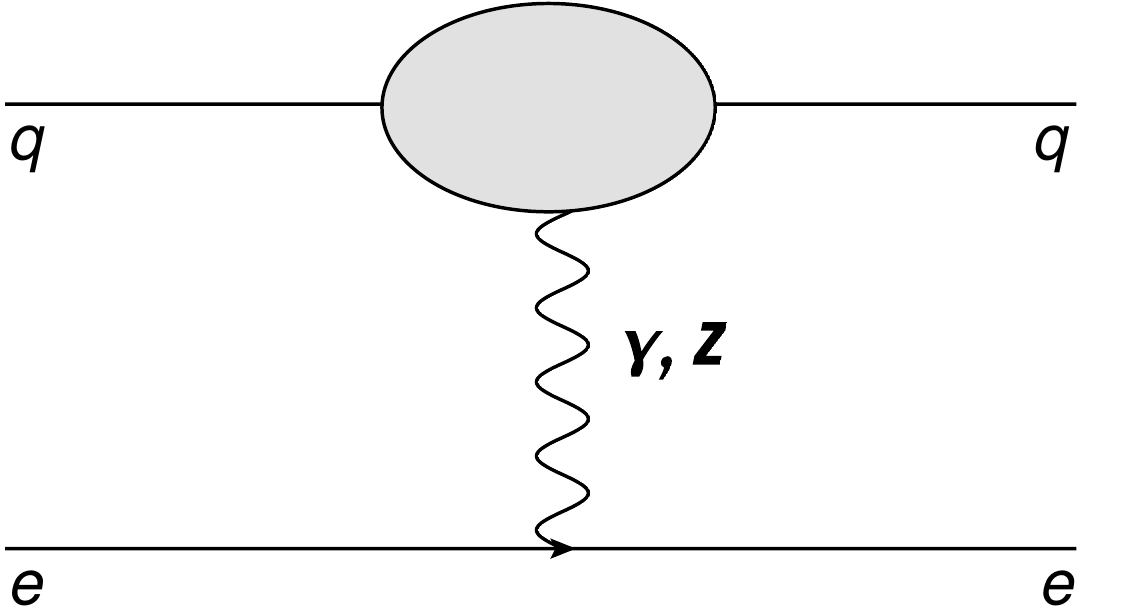}}
\end{minipage}
\caption{Classes of corrections to the $eq$ scattering due to a leptophobic $Z'$: $\gamma-Z'$ (or $Z-Z'$) mixing and vertex corrections in the quark side of the diagram. The external leg (quark self-energy) corrections are not shown.}
\label{fig:classes}
\end{figure}

A leptophobic $Z'$ with vector and axial-vector couplings to quarks will in general modify the value of the $C_{iq}$ coefficients, and thus it will have measurable consequences on the PV $eD$ asymmetry. Due to its leptophobic nature the modifications can be classified as vertex and external leg (quark propagator) corrections in the quark side of the diagram and $\gamma$ or $Z$ mixing with the new neutral gauge boson $Z'$, as shown diagrammatically in Fig.~\ref{fig:classes}. The latter effect was studied in a recent publication \cite{Buckley:2012tc} that we will complement in this work by analyzing the effect of the vertex correction and external leg correction (not shown).


We will follow a phenomenological, bottom-up approach, where the $Z'$ mass and its couplings to quarks are free parameters. Specifically, we will use the following Lagrangian to describe the interaction of the leptophobic $Z'$ boson with quarks
\begin{eqnarray}
\label{Z'lagrangian}
{\cal L} (x) = g' Z'_\mu \sum_{ij} \bar{q}_i \gamma^\mu \left( Q_{ij}^{'V} + Q_{ij}^{'A} \gamma_5 \right) q_j~.
\end{eqnarray}
where $g'$ is the new gauge coupling constant and $Q_{ij}^{'\scriptstyle{V(A)}}$ are the flavor-dependent vector (axial-vector) quark couplings, that satisfy $Q_{ij}^{'\scriptstyle{V(A)}\star} = Q_{ji}^{'\scriptstyle{V(A)}}$ to ensure the hermiticity of the Lagrangian. In this expression the quark fields are the mass eigenstates, i.e. the CKM rotation has already been performed.

\begin{figure}[h!]
\includegraphics[width=4.5cm]{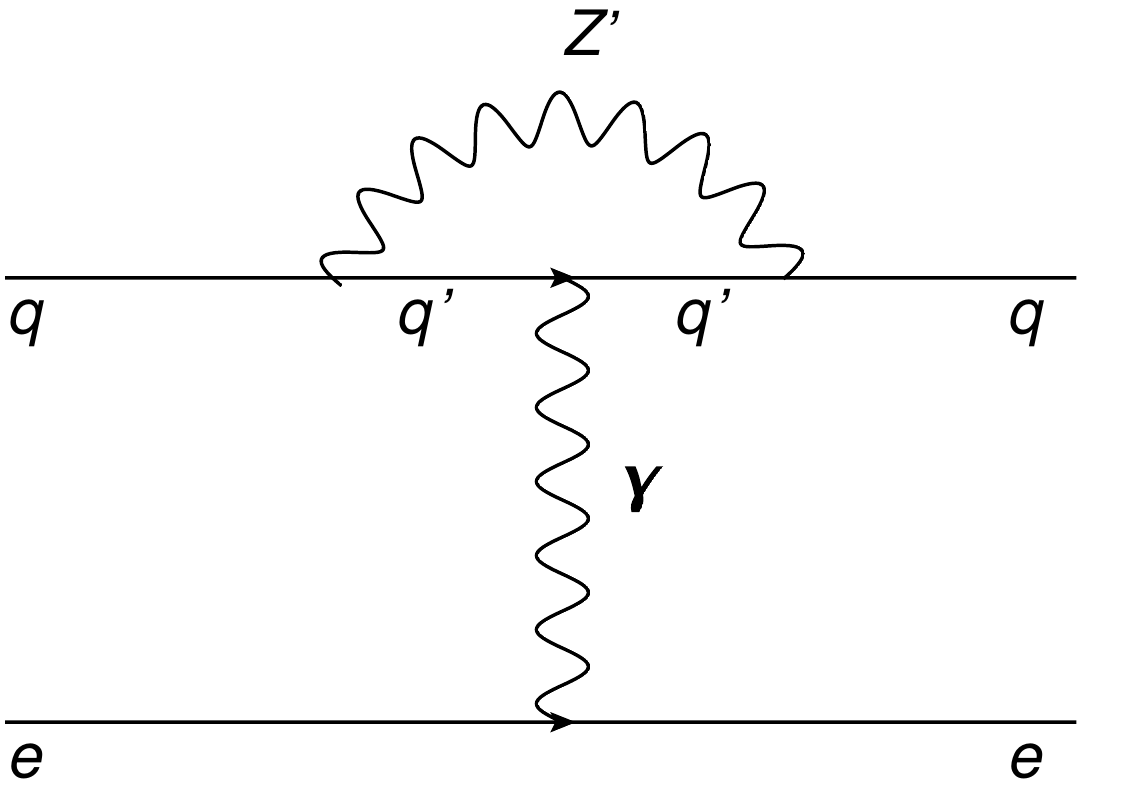}
\caption{New penguin diagram built with the $Z'$ correction to the $\gamma\bar{q}q$ vertex, that generates a non-zero BSM contribution to $C_{2q}$.}
\label{fig:ChargeRadii}
\end{figure}

Both the $Z$ and $\gamma$ couplings to quarks receive corrections from one-loop diagrams containing the new $Z'$ boson. In turn, these vertex corrections (and the related external leg corrections) generate new contributions to the PV $C_{iq}$ coefficients through penguin-like diagrams, as shown in Fig.~\ref{fig:ChargeRadii}. Although we have of course included the external leg corrections in our computation, we have omitted them from Fig.~\ref{fig:ChargeRadii} for the sake of brevity.

If the $Z'$ is much heavier than the quark $j$ running in the loop, we can neglect terms of order $m^2_{j}/M_{Z'}^2$. In this case, the sum of vertex and external leg corrections to both the $\gamma {\bar q}q$ and $Z {\bar q} q$ interactions vanish in the $Q^2\to 0$ limit. The leading non-vanishing correction that transforms as an axial vector coupling occurs at second order in the four-momentum transfer. For the $Z{\bar q}q$ interaction, the resulting contribution to the four-fermion $({\bar e}\gamma^\mu e)\, ({\bar q}\gamma_\mu\gamma_5 q)$ operator is suppressed by $Q^2/M_Z^2$, making it negligible at the kinematics of the future PV-DIS experiments. In contrast, the ${\cal O}(Q^2)$ contribution to the axial vector $\gamma{\bar q} q$ vertex, commonly referred to as the \lq\lq anapole moment"\cite{Musolf:1990sa}, compensates for the $1/Q^2$ arising from the photon propagator in the penguin graph of Fig.~\ref{fig:ChargeRadii}, leading to the following $Q^2$-independent contribution to the $C_{iq}$: 
\begin{eqnarray}
\label{eq:C1qgamma}
\delta C_{1q}		& = & 0~,\\
\delta C_{2q}		& = & \frac{-2\,Q_q}{3\pi^2} g^{\prime 2} s_{\mathrm{W}}^2 c_{\mathrm{W}}^2 ~\frac{M_Z^2}{M_{Z'}^2} Q_{qj}^{'V} Q_{qj}^{'A} \left(\!\frac{1}{6} - \log{\frac{m_j^2}{M_{Z'}^2}} \!\right),\nonumber
\label{eq:C2qZ}
\end{eqnarray}
where $c_{\mathrm{W}}$ and $s_{\mathrm{W}}$ are the cosine and sine of the weak mixing angle $\theta_{\mathrm{W}}$. In this expression and in the remainder of this article we assume the couplings $Q_{ij}^{'\scriptstyle{V(A)}}$ to be real for the sake of simplicity. Since the photon does not have any axial-vector coupling, the $C_{1q}$ coefficients do not receive any BSM contribution. On the other hand, the $Z'$ contribution to the anapole moment of the quarks generates the above-given non-zero $\delta C_{2q}$.

It is worth stressing that these results are model-independent, in the sense that they do not depend on the details of the UV completion of the $Z'$ model but only on its mass and its couplings to quarks.

One might be interested in the same expression for the case when the mass of the quark running in the loop is not negligible compared with the $Z'$ mass, as would happen with a ${\cal O}(100)$ GeV $Z'$ particle with non-diagonal couplings to up and top quarks. In this case, however, this result depends on the UV completion of the theory and hence a model-independent evaluation is no longer possible. In particular, the $Z'$ contribution to the $Z$ or $\gamma$ vertex correction shows a gauge dependence that will be cancelled by other new contributions in the more complete theory where the $Z'$ is embedded\footnote{Notice that this situation is the same in the Standard Model if we substitute the $Z'$ by the electroweak $Z$ boson. In that case the gauge dependence is canceled by one-loop diagrams with the scalar degrees of freedom coming from the Higgs sector.}. This issue was also discussed in Ref.~\cite{Gresham:2012wc}, where the ${\cal O} (m_t^2 / M_{Z'}^2)$ effect on the $C_{1q}$ coefficient due to a light $Z'$ boson with non-diagonal flavor coupling to up and top quarks was studied in a specific model where the new neutral gauge bosons are introduced through a non-Abelian flavor gauge symmetry \cite{Jung:2011zv}. Models where the $Z'$ has a fairly large non-diagonal coupling between up and top quarks have received some attention recently \cite{ZprimeForTop} since they could accommodate the large measured top quark forward-backward asymmetry $A_{FB}^t$ at the Tevatron \cite{TevatronAsymmetry}. 
 
It is useful to notice at this point that a $Z'$ boson with up-charm, down-strange or down-bottom flavor changing couplings would contribute at tree-level to the $D^0-\bar{D}^0$, $K^0-\bar{K}^0$ or $B_d^0-\bar{B}^0_d$ mass splittings respectively. Namely, if we assume small $Z-Z'$ mixing we find \cite{Langacker:2000ju}
\begin{eqnarray}
\label{eq:DeltaM}
\frac{\left[ \Delta m_P \right]_{Z'}}{m_P}
	&\approx& \frac{1}{3} \left( \frac{g' F_P}{M_{Z'}} \right)^2  \times \\
	&\times& \left[ \left(Q_{ij}^{'A}\right)^2 (k_P +2) - \left(Q_{ij}^{'V}\right)^2 (k_P -2) \right]~,\nonumber
\end{eqnarray}
where $m_P$ and $F_P$ are the mass and decay constants of the $P$ meson ($P=D,K,B_d$);  $i,j$ denote the valence quarks in $P$; and $k_P$ is given by
\begin{eqnarray}
k_P \equiv \frac{3}{2} + \frac{m^2_P}{(m_i+m_j)^2}~.
\end{eqnarray}
Thus, we see that unless an unnatural cancellation between the V and A terms in Eq.~\eqref{eq:DeltaM} occurs, this result can be used to set strong bounds on these flavor non-diagonal couplings. 

It has been shown experimentally that these three mass splittings $\Delta m_{B,D,K}$ are non-zero \cite{Beringer:1900zz}. These measurements are actually very precise in the  $B$ and $K$ systems, where the relative errors are at the 0.1-1.0\% level. However, the determination of the mass splittings in the Standard Model is quite complicated, especially in the $D$ and $K$ cases due to long-distance effects, which opens the possibility for large NP contribution. Nonetheless, even in the extreme scenario where the BSM contribution to $\Delta m_P$ is of the same size of the experimental value, we still find that
\begin{eqnarray}
g' |Q_{ij}^{'V,A} | \frac{M_Z}{M_{Z'}} \ll 10^{-3}~,
\end{eqnarray}
for the three cases $(i,j)=(u,c),(d,s),(d,b)$ \footnote{For a more detailed analysis of the effect of a non-universal $Z'$ in meson-antimeson mixing see Ref.~\cite{He:2007iu} and reference therein.}. Such a strong constraint makes their effect on the PV coefficients $C_{2q}$ completely negligible:
\begin{eqnarray}
|\delta C_{2q}|	 ~\ll~ |Q_q| ~ 10^{-8}~ \left| \frac{1}{6} - \log{\frac{m_j^2}{M_{Z'}^2}} \right|~.
\end{eqnarray}

We can therefore safely neglect the flavor non-diagonal terms in our $C_{iq}$ expressions, except for the up-top coupling that we will keep. After some trivial manipulations we can then write
\begin{eqnarray}
\label{eq:C1d-vertex}
\delta C_{1d}		& = & 0~,\\
\delta C_{1u}		& = & B^{(1)}_{ut}~,\\
\label{eq:C2d-vertex}
\delta C_{2d}		& \approx &	\frac{4}{9\pi^2}  g^{\prime 2} s_{\mathrm{W}}^2 c_{\mathrm{W}}^2 \frac{M_Z^2}{M_{Z'}^2}Q_{dd}^{'V} Q_{dd}^{'A} \times \nonumber\\
				& &			\times \left( 4.6 + \log{\frac{M_{Z'}}{M_Z}} \right) \\
\label{eq:C2u-vertex}
\delta C_{2u}		& \approx &	\frac{-8}{9\pi^2}  g^{\prime 2}  s_{\mathrm{W}}^2 c_{\mathrm{W}}^2 \frac{M_Z^2}{M_{Z'}^2} Q_{uu}^{'V} Q_{uu}^{'A}\times  \nonumber\\
					&&		\times\left( 4.6 + \log{\frac{M_{Z'}}{M_Z}} \right) + B^{(2)}_{ut}~,
\end{eqnarray}
where we have replaced the light quark masses $m_{u,d}$ appearing in Eq.~\eqref{eq:C2qZ} by a hadronic scale of $1$ GeV, as a conservative cut-off, since non-perturbative effects become large for smaller energies. The $B^{(1,2)}_{ut}$ are model-dependent quantities with the following asymptotic behavior in the limit $M_{Z'}\gg m_t$:
\begin{eqnarray}
B^{(1)}_{ut}		\!\!& \to &\!\! 0~,\\
B^{(2)}_{ut}		\!\!& \to &\!\! \frac{-4}{9\pi^2} g^{\prime 2} s_{\mathrm{W}}^2 c_{\mathrm{W}}^2 Q_{ut}^{'V} Q_{ut}^{'A} \frac{M_Z^2}{M_{Z'}^2} \!\left(\! \frac{1}{6} - \log{\frac{m^2_t}{M^2_{Z'}}} \!\right)\!\!.
\end{eqnarray}

As shown in Fig.~\ref{fig:classes}, a leptophobic $Z'$ also modifies the $C_{2q}$ coefficients through photon-$Z'$ mixing, although it cannot affect $C_{1q}$ as the photon does not have the tree-level axial coupling to the electron needed to generate the $({\bar e}\gamma^\mu \gamma_5 e)\, ({\bar q}\gamma_\mu q)$ operator. On the other hand the $Z-Z'$ mixing diagram does contribute to both $C_{1q}$ and $C_{2q}$ coefficients, but the associated mixing angle is constrained to be so small \cite{Erler:2009ut} that this contribution can be safely neglected.

The $\gamma$-$Z'$ correction to $C_{2q}$ was calculated recently with the following result \cite{Buckley:2012tc}
\begin{eqnarray}
\label{eq:propagator}
\delta C_{2q} & \approx & \frac{8}{9\pi^2}  g^{\prime 2} s_{\mathrm{W}}^2 c_{\mathrm{W}}^2 \left(\frac{M_Z}{M_{Z'}}\right)^2 Q^{'A}_{qq} \times  \label{eq:deltaC} \\
& & \left[ 10.54~(2Q^{'V}_{uu}-Q^{'V}_{dd}) - 8.65~Q^{'V}_{ss}+ \right. \nonumber \\
 & & \left. 12.43~Q^{'V}_{cc} - 4.44~Q^{'V}_{bb} - 1.89~Q^{'V}_{tt}\right]~, \nonumber  
\end{eqnarray}
where we have conveniently chosen a form similar to Eqs.~\eqref{eq:C2d-vertex} and \eqref{eq:C2u-vertex} to ease the comparison.

We observe a complementarity between both contributions to $C_{2q}$ (vertex and propagator corrections):
\begin{itemize}
\item Non-diagonal flavor couplings do not contribute to the $\gamma-Z'$ (or $Z-Z'$) mixing since the SM neutral current couplings are flavor diagonal, but they do contribute to the vertex corrections. This is particularly relevant for a $Z'$ that couples to up and top quarks;
\item The dependence on the $Z'$ mass is different, due to the presence of a logarithmic term in the vertex correction. However, we wrote Eqs.~\eqref{eq:C2d-vertex} and \eqref{eq:C2u-vertex} in such a way that the logarithm contribution is subdominant for $Z'$ masses in the 100-1000 GeV range;
\end{itemize}

Moreover, in the case of flavor diagonal couplings:
\begin{itemize}
\item Obviously the corrections to $C_{2q}$ are zero unless the $Z'$ has axial-vector couplings to the $\bar{q}q$ pair ($Q^{'A}_{qq}$). However, for the vertex correction we need also a non-zero vector coupling to the same $\bar{q}q$ pair ($Q^{'V}_{qq}$), whereas in the $\gamma-Z'$ mixing the vector coupling can be to a different quark pair ($Q^{'V}_{q'q'}$);
\item The $\gamma-Z'$ mixing correction is generally larger in magnitude than the vertex correction, unless some cancellation between the different terms in Eq.~\eqref{eq:propagator} takes place. 
\end{itemize}

Therefore, our final result for the $C_{iq}$ coefficients is given by the sum of the vertex corrections of Eqs.~\eqref{eq:C1d-vertex}-\eqref{eq:C2u-vertex} and $\gamma-Z'$ mixing correction given in Eq.~\eqref{eq:propagator}. 
In particular, if the $Z'$ couplings are flavor universal and diagonal these expressions take the following form
\begin{eqnarray}
\delta C_{2d} & \approx & \frac{4}{9\pi^2}  g^{\prime 2} s_{\mathrm{W}}^2 c_{\mathrm{W}}^2 \left(\frac{M_Z}{M_{Z'}}\right)^2 Q^{'A}_{dd} \\
& & \!\!\!\!\!\!\!\!\!\! \times \left[ 63.24~ Q^{'V}_{uu}  + Q^{'V}_{dd} \left( -42.66 + \log{\frac{M_{Z'}}{M_Z}} \right) \right]~, \nonumber  \\
\delta C_{2u} & \approx & \frac{-8}{9\pi^2}  g^{\prime 2} s_{\mathrm{W}}^2 c_{\mathrm{W}}^2 \left(\frac{M_Z}{M_{Z'}}\right)^2 Q^{'A}_{uu} \\
& & \!\!\!\!\!\!\!\!\!\! \times \left[ Q^{'V}_{uu} \left( -26.92 + \log{\frac{M_{Z'}}{M_Z}} \right) + 23.63~ Q^{'V}_{dd} \right] \nonumber
\end{eqnarray}

For the sake of illustration, we now apply our results to two specific $Z'$ scenarios that have been studied recently in the literature as possible explanations for the $W^\pm+jj$ excess observed by CDF Collaboration in the study of events with a lepton, missing transverse energy and a pair of hadronic jets \cite{Aaltonen:2011mk}

The first case is given by the leptophobic $E_6$ GUT scenario outlined in Ref.~\cite{Barger:1996kr}, where the $Z'$ couplings to quarks are flavor diagonal and universal, with $Q_{uu}^{'V} = 1/6$, $Q_{uu}^{'A} = 1/2$, $Q_{dd}^{'V} =  -1/3$ and $Q_{dd}^{'A} = 0$. In this model the gauge coupling $g'$ and the mass of the $Z'$ boson are not fixed, but it is illustrative to use $g'\sim 0.6$ and $M_{Z'} \sim 150$~GeV, since these values can accommodate the CDF $W^\pm+jj$ excess, as shown in Ref.~\cite{Buckley:2011mm}. Hence we find in this benchmark scenario at $Q^2=0$ the following result
\begin{eqnarray}
\delta C_{2u} & = &  +0.0131 \left(\frac{150~\mbox{GeV}}{M_{Z'}}\right)^2\left(\frac{g'}{0.6}\right)^2 \nonumber\\
						& &	\times \left(  1-0.014 \log{\frac{M_{Z'}}{150~\mbox{GeV}}}  \right), \label{eq:C2uE6} \\
\delta C_{2d} & = & 0. \label{eq:C2dE6}
\end{eqnarray}
This correction to the $C_{2u}$ coefficient, representing approximately a $30\%$ correction to the SM value, is completely dominated by the $Z'-\gamma$ mixing term, and thus the conclusions of Ref.~\cite{Buckley:2012tc} are not substantially modified by the inclusion of the vertex correction that only decreases the result by $\sim 7\%$ \footnote{The overall sign of the correction $\delta C_{2u}$ is however the opposite than in Ref.~\cite{Buckley:2012tc}, due to a typo in the sign of the axial-vector coupling of up quarks $Q_{uu}^{'A}$ in that work.}. 
It must be emphasized, however, that this is not true in general and one could have $Z'$ models were the vertex correction is the dominant. Nevertheless these cases are less interesting phenomenologically, since the value of $\delta C_{2u}$ is then smaller. Likewise the absence of a $C_{2d}$ correction is due to the zero value of the down quarks axial-vector charge ($Q_{dd}^{'A}$) in our benchmark $Z'$ scenario and hence is not a model-independent feature. 

We note the results shown so far have been obtained at $Q^2=0$, whereas the future PVDIS experiments will take place at a finite value of $Q^2$. In the case of the SOLID experiment the $Q^2$ will be in the range 4 - 10 (GeV/c)$^2$. In Ref.~\cite{Buckley:2012tc} it was found that this finite-$Q^2$ effect reduces the absolute value of $\delta C_{2u}$ by $\sim 25 \%$ ($\sim 30 \%$) at the lower (upper) end of the kinematic range, a modification that can also be applied also to our results for this first example since the final result is dominated by the propagator corrections. 

A deviation of this size ($2C_{2u}-C_{2d} \sim 0.02$) from the Standard Model prediction will be accessible to SOLID and the EIC, therefore probing this class of $Z'$ models and evaluating the presence of axial-vector couplings to quarks. 

As a second example, we discussed now the phenomenological model introduced in Ref.~\cite{Buckley:2011vc} to explain the CDF $W^\pm+jj$ excess \cite{Aaltonen:2011mk}, through a light leptophobic $Z'$ that couples only to first generation quarks. 

\begin{figure}[h!]
\vspace{0.6cm}
\begin{minipage}[t]{.7\linewidth}\centering
\centerline{\includegraphics[width=8.5cm]{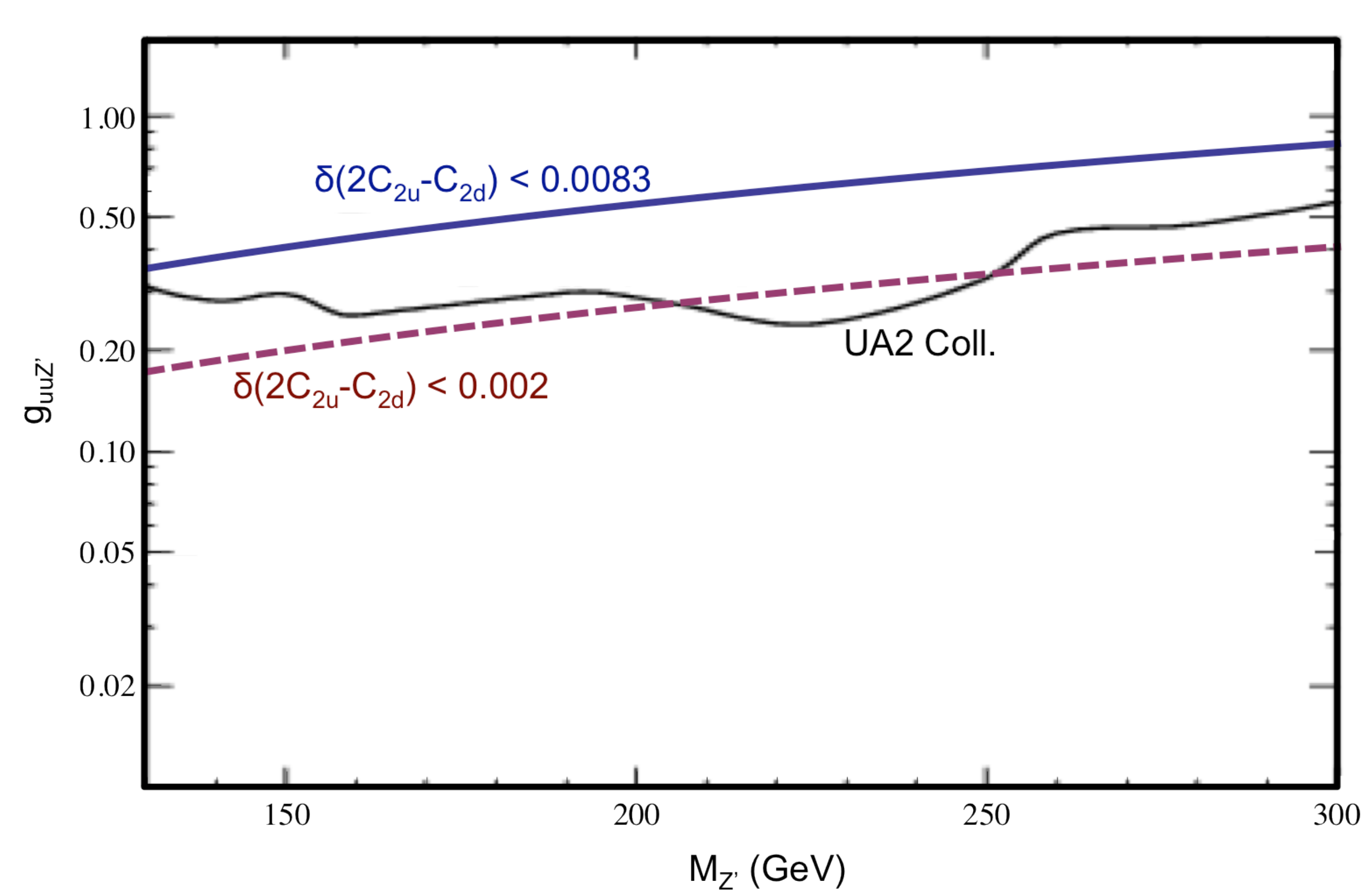}}
\end{minipage}
\vspace{1cm}
\begin{minipage}[t]{.7\linewidth}\centering
\centerline{\includegraphics[width=8.5cm]{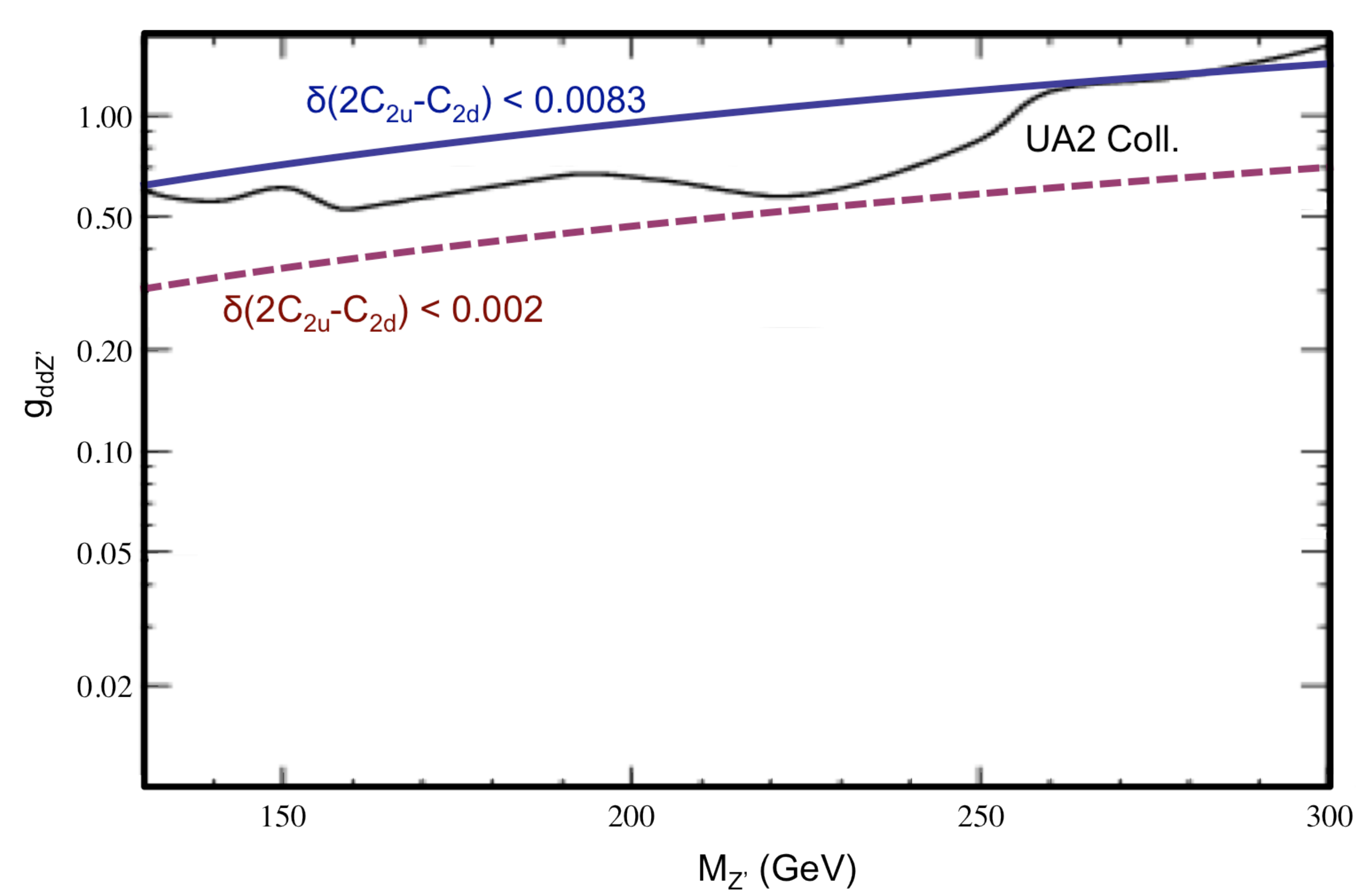}}
\end{minipage}
\caption{Regions of the $g_{qqZ'}-M_{Z'}$ plane probed by the UA2 experiment as extracted in Ref.~\cite{Buckley:2011vc}, and by future measurements of the quantity $2C_{2u}-C_{2d}$ with different levels of precision. The upper and lower plots show the result for a $Z'$ that only couples to $u$ and $d$ quarks respectively.}
\label{fig:UA2comparison}
\end{figure}

As shown in Ref.~\cite{Buckley:2011vc}, the collider bounds on such a $Z'$ are not strong, and in fact for a light $Z'$ (below 300 GeV) the strongest bounds come from the dijet searches at the UA2 experiment \cite{Alitti:1993pn} ($p\bar{p} \to Z' \to q\bar{q}$). At the Tevatron, in contrast, the QCD background become too large compared with the signal. The comparison of these bounds, extracted in Ref.~\cite{Buckley:2011vc}, and the bounds that can be derived from the measurements of the PV $eeqq$ coefficients $C_{2q}$ are shown in Fig.~\ref{fig:UA2comparison}. As we can see, a future determination of the quantity $2C_{2u}-C_{2d}$ with an error in the $10^{-2}-10^{-3}$ range would probe a comparable portion of the parameter space as the UA2 measurements. To facilitate the comparison, we follow in this plot the notation of Ref.~\cite{Buckley:2011vc}, where the $Z'$ is assume to couple either to left- or right-handed quarks, but not both at the same time, and where their coupling constant $g_{qqZ'}$ is related to our coupling constants as follows
\begin{eqnarray}
g' Q^{V \prime}_{qq} = \frac{g_{qqZ'}}{2}~,~~~~~~ g' Q^{A \prime}_{qq} = \pm \frac{g_{qqZ'}}{2}\ \ \ .
\end{eqnarray}
Here, the plus (minus) sign corresponds to the case when the $Z'$ boson couples to right-handed (left-handed) quarks.

In particular, it was shown in Ref.~\cite{Buckley:2011vc} that a $Z'$ with a mass of 150 GeV and couplings to left-handed first-generation quarks could explain the CDF $W^\pm+jj$ excess. Applying our results, this $Z'$ would produce the following correction to $2C_{2u}-C_{2d}$
\begin{eqnarray}
\delta \left( 2C_{2u}-C_{2d} \right) 
&=& -0.012 \left[ 4g_{uuZ'}^2 + g_{ddZ'}^2 \right. \nonumber\\
&& \left. - 5.3~ g_{uuZ'} g_{ddZ'} \right]~.
\end{eqnarray}
As shown in Fig. 4 of Ref.~\cite{Buckley:2011vc}, the values of the couplings $g_{uuZ'}$ and $g_{ddZ'}$ needed to explain the $W^\pm+jj$ excess are not fixed separately. However we can use the point $(g_{uuZ'},g_{ddZ'}) \sim (0.25,0.00)$ as an illustrative example, and in that case we find $\delta \left( 2C_{2u}-C_{2d} \right) \approx -0.003$, an effect smaller than in the $E_6$ GUT scenario, but still above the permil level. Probing an effect at this level would likely require the sensitivity of a future experiment, as one might envision at an EIC.

In the same work \cite{Buckley:2011vc} it is argued that one could also explain the large measured top quark forward-backward asymmetry $A_{FB}^t$ at the Tevatron \cite{TevatronAsymmetry}, if the $Z'$ has a fairly large non-diagonal coupling between up and top quarks. These particles would certainly generate additional contributions to the  $C_{1u}$ and $C_{2u}$ coefficients, but as we have observed this contributions have to examined on a model by model basis, since their new $Z^\prime$ bosons are not much heavier than the top quark.

To summarize, we have shown how the future measurement of $C_{1q}$ and $C_{2q}$ will be sensitive to several couplings of a leptophobic $Z'$, and will provide interesting information about the flavor structure of these BSM models. As an example, we see that a disagreement of future measurements with the $C_{1u}$ SM prediction could only be due by a $Z'tu$ coupling, whereas a disagreement in $C_{1d}$ could hardly be explained by leptophobic $Z'$. On the other hand, a disagreement only in $C_{2u}$ could probe some of the leptophobic $Z'$ models proposed to explain the Tevatron W+dijet anomaly. It is important to emphasize that the corrections to the $C_{2q}$ coefficients due to flavor diagonal couplings vanish if the $Z'$ does not have axial-vector couplings to the $q\bar{q}$ pair. For this reason, models with only vector couplings, like the one studied in Ref.~\cite{Yu:2011cw}, cannot be probed with PV DIS electron-deuteron scattering.

More generally, the leptophobic $Z'$ models studied in this paper, wherein the coefficients $C_{1q}$ might not receive any significant correction, show nicely how future $C_{2q}$ determinations via $A_{PV}^{eD}$ are an interesting BSM probe, complementary to the precise $C_{1q}$ determinations through atomic PV and PV elastic $ep$ scattering.

\noindent{\it  Acknowledgements} 
We thank Sean Tulin and Martin Jung for useful discussions and Sonny Mantry and John Ng for helpful conversations in formulating the initial phase of this analysis. This work was supported in part by DOE contract DE-FG02-08ER41531 and the Wisconsin Alumni Research Foundation.


\begin{thebibliography}{99}  


  
\bibitem{Yu:2011cw} 
  F.~Yu,
  Phys.\ Rev.\ D {\bf 83}, 094028 (2011)
  [arXiv:1104.0243 [hep-ph]].
  
\bibitem{Buckley:2011mm} 
  M.~R.~Buckley, D.~Hooper and J.~L.~Rosner,
  Phys.\ Lett.\ B {\bf 703}, 343 (2011)
  [arXiv:1106.3583 [hep-ph]].

\bibitem{Buckley:2011vc}
  M.~R.~Buckley, D.~Hooper, J.~Kopp, E.~Neil,
  Phys.\ Rev.\  {\bf D83}, 115013 (2011).
  [arXiv:1103.6035 [hep-ph]].  

\bibitem{Buckley:2011vs} 
  M.~Buckley, P.~Fileviez Perez, D.~Hooper and E.~Neil,
  Phys.\ Lett.\ B {\bf 702}, 256 (2011)
  [arXiv:1104.3145 [hep-ph]].

\bibitem{Cheung:2011zt} 
  K.~Cheung and J.~Song,
  Phys.\ Rev.\ Lett.\  {\bf 106}, 211803 (2011)
  [arXiv:1104.1375 [hep-ph]].
 
\bibitem{Subedi:2011zz} 
  R.~R.~Subedi, X.~Deng, R.~Michaels, K.~Pan, P.~E.~Reimer, D.~Wang and X.~Zheng,
  AIP Conf.\ Proc.\  {\bf 1374}, 602 (2011).

\bibitem{Dudek:2012vr} 
  J.~Dudek, R.~Ent, R.~Essig, K.~Kumar, C.~Meyer, R.~McKeown, Z.~E.~Meziani and G.~A.~Miller {\it et al.},
  arXiv:1208.1244 [hep-ex].
  
\bibitem{SOLID}
The SoLID Experiment, Jefferson Laboratory Experiment E12-10-007, P.A. Souder (contact person), 
http://www.jlab.org/$\mathrm{exp}_{-}$prog/PACpage/PAC37/ proposals/Proposals/Previously$\%$20Approved/E12-10-007.pdf

\bibitem{Boer:2011fh} 
  D.~Boer, M.~Diehl, R.~Milner, R.~Venugopalan, W.~Vogelsang, D.~Kaplan, H.~Montgomery and S.~Vigdor {\it et al.},
  arXiv:1108.1713 [nucl-th].
      
\bibitem{Marciano:1982mm} 
  W.~J.~Marciano and A.~Sirlin,
  Phys.\ Rev.\ D {\bf 27}, 552 (1983).
  
\bibitem{Musolf:1990ts} 
  M.~J.~Musolf and B.~R.~Holstein,
  Phys.\ Lett.\ B {\bf 242}, 461 (1990).

\bibitem{Beringer:1900zz}
  J.~Beringer {\it et al.}  [Particle Data Group Collaboration],
  Phys.\ Rev.\ D {\bf 86} (2012) 010001.
    
        
\bibitem{Young:2007zs} 
  R.~D.~Young, R.~D.~Carlini, A.~W.~Thomas and J.~Roche,
  Phys.\ Rev.\ Lett.\  {\bf 99}, 122003 (2007)
  [arXiv:0704.2618 [hep-ph]].
     
\bibitem{Wood:1997zq} 
  C.~S.~Wood, S.~C.~Bennett, D.~Cho, B.~P.~Masterson, J.~L.~Roberts, C.~E.~Tanner and C.~E.~Wieman,
  Science {\bf 275}, 1759 (1997).
  
\bibitem{Porsev:2009pr} 
  S.~G.~Porsev, K.~Beloy and A.~Derevianko,
  Phys.\ Rev.\ Lett.\  {\bf 102}, 181601 (2009)
  [arXiv:0902.0335 [hep-ph]].

\bibitem{QWeak} http://www.jlab.org/qweak/


\bibitem{Mantry:2010ki} 
  S.~Mantry, M.~J.~Ramsey-Musolf and G.~F.~Sacco,
  Phys.\ Rev.\ C {\bf 82}, 065205 (2010)
  [arXiv:1004.3307 [hep-ph]].

\bibitem{KK} K. Kumar, private communication. 

\bibitem{Musolf:1990sa} 
  M.~J.~Musolf and B.~R.~Holstein,
  Phys.\ Rev.\ D {\bf 43}, 2956 (1991).


\bibitem{Buckley:2012tc} 
  M.~R.~Buckley and M.~J.~Ramsey-Musolf,
  Phys.\ Lett.\ B {\bf 712}, 261 (2012)
  [arXiv:1203.1102 [hep-ph]].
  
\bibitem{Gresham:2012wc} 
  M.~I.~Gresham, I.~-W.~Kim, S.~Tulin and K.~M.~Zurek,
  arXiv:1203.1320 [hep-ph].

\bibitem{Jung:2011zv} 
  S.~Jung, A.~Pierce and J.~D.~Wells,
  Phys.\ Rev.\ D {\bf 83}, 114039 (2011)
  [arXiv:1103.4835 [hep-ph]].

\bibitem{ZprimeForTop} 
  S.~Jung, H.~Murayama, A.~Pierce and J.~D.~Wells,
  Phys.\ Rev.\ D {\bf 81}, 015004 (2010)
  [arXiv:0907.4112 [hep-ph]].
  Q.~-H.~Cao, D.~McKeen, J.~L.~Rosner, G.~Shaughnessy and C.~E.~M.~Wagner,
  Phys.\ Rev.\ D {\bf 81}, 114004 (2010)
  [arXiv:1003.3461 [hep-ph]].
  M.~I.~Gresham, I.~-W.~Kim and K.~M.~Zurek,
  Phys.\ Rev.\ D {\bf 84}, 034025 (2011)
  [arXiv:1102.0018 [hep-ph]].
  J.~Cao, L.~Wang, L.~Wu and J.~M.~Yang,
  Phys.\ Rev.\ D {\bf 84}, 074001 (2011)
  [arXiv:1101.4456 [hep-ph]].
  V.~Barger, W.~-Y.~Keung and C.~-T.~Yu,
  Phys.\ Lett.\ B {\bf 698}, 243 (2011)
  [arXiv:1102.0279 [hep-ph]].
  B.~Bhattacherjee, S.~S.~Biswal and D.~Ghosh,
  Phys.\ Rev.\ D {\bf 83}, 091501 (2011)
  [arXiv:1102.0545 [hep-ph]].
  E.~L.~Berger, Q.~-H.~Cao, C.~-R.~Chen, C.~S.~Li and H.~Zhang,
  Phys.\ Rev.\ Lett.\  {\bf 106}, 201801 (2011)
  [arXiv:1101.5625 [hep-ph]].

\bibitem{TevatronAsymmetry} 
  V.~M.~Abazov {\it et al.}  [D0 Collaboration],
  Phys.\ Rev.\ Lett.\  {\bf 100}, 142002 (2008)
  [arXiv:0712.0851 [hep-ex]].
  T.~Aaltonen {\it et al.}  [CDF Collaboration],
  Phys.\ Rev.\ D {\bf 83}, 112003 (2011)
  [arXiv:1101.0034 [hep-ex]].

\bibitem{Langacker:2000ju} 
  P.~Langacker and M.~Plumacher,
  Phys.\ Rev.\ D {\bf 62}, 013006 (2000)
  [hep-ph/0001204].

\bibitem{He:2007iu} 
  X.~-G.~He and G.~Valencia,
  Phys.\ Lett.\ B {\bf 651}, 135 (2007)
  [hep-ph/0703270];
%
  A.~J.~Buras, F.~De Fazio and J.~Girrbach,
  arXiv:1211.1896 [hep-ph].
      
\bibitem{Erler:2009ut} 
  J.~Erler, P.~Langacker, S.~Munir and E.~Rojas,
  AIP Conf.\ Proc.\  {\bf 1200}, 790 (2010)
  [arXiv:0910.0269 [hep-ph]].

\bibitem{Barger:1996kr}
  V.~D.~Barger, K.~M.~Cheung, P.~Langacker,
  Phys.\ Lett.\  {\bf B381}, 226-236 (1996).
  [hep-ph/9604298].
  K.~S.~Babu, C.~F.~Kolda and J.~March-Russell,
  Phys.\ Rev.\  D {\bf 54}, 4635 (1996)
  [arXiv:hep-ph/9603212].
 C.~F.~Kolda,
  Nucl.\ Phys.\ Proc.\ Suppl.\  {\bf 52A}, 120-126 (1997).
  [hep-ph/9606396];
  J.~L.~Rosner,
  Phys.\ Lett.\  {\bf B387}, 113-117 (1996).
  [hep-ph/9607207];
  V.~D.~Barger, N.~G.~Deshpande, K.~Whisnant,
  Phys.\ Rev.\ Lett.\  {\bf 56}, 30 (1986);
 V.~D.~Barger, N.~G.~Deshpande, J.~L.~Rosner, K.~Whisnant,
  Phys.\ Rev.\  {\bf D35}, 2893 (1987).

\bibitem{Aaltonen:2011mk}
  T.~Aaltonen {\it et al.} [CDF Collaboration],
  Phys.\ Rev.\ Lett.\  {\bf 106}, 171801 (2011).
  [arXiv:1104.0699 [hep-ex]].

\bibitem{Alitti:1993pn}
  J.~Alitti {\it et al.} [UA2 Collaboration],
  Nucl.\ Phys.\  {\bf B400}, 3-24 (1993).  
  
\end{thebibliography}
\end{document}